\documentclass[a4paper,11pt]{article}
\usepackage{jinstpub}
\usepackage{bm}
\title{\boldmath Study of active geomagnetic shielding coils system for JUNO}
\author[a,1]{G.~Zhang,}
\author[b,1]{J.~Songwadhana,}
\author[a,1]{H.~Lu,}
\author[b,1]{Y.~Yan,}
\author[c]{N.~Morozov,}
\author[a]{F.~Ning,}
\author[a]{P.~Zhang,}
\author[a]{C.~Yang,}
\author[b]{K.~Khosonthongkee,}
\author[b]{A.~Limphirat,}
\author[b]{T.~Yan,}
\author[e]{T.~Payupol,}
\author[e]{N.~Suwonjandee,}
\author[e]{B.~Asavapibhop,}
\author[f]{U.~Sawangwit,}
\author[g]{A.~Sangka,}
\author[a]{Z.~Zhu,}
\author[a,d]{X.~Wang,}
\author[a,d]{X.~Liu}
\author[a,d]{and Z.~Xie}
\affiliation[a]{Institute of High Energy Physics, Chinese Academy of Sciences\\
Beijing 100049, China}
\affiliation[b]{School of Physics, Suranaree University of Technology,\\
111 University Avenue, Nakhon Ratchasima, Thailand}
\affiliation[c]{Joint Institute for Nuclear Research (JINR), Joliot-Curie 6, Dubna,\\
Moscow Oblast, Russian Federation}
\affiliation[d]{University of Chinese Academy of Sciences,\\
Beijing 100049, China}
\affiliation[e]{Particle Physics Research Laboratory, Department of Physics, Faculty of Science, Chulalongkorn University \\
254 Phayathai Rd., Patumwan, Bangkok 10330 Thailand}
\affiliation[f]{National Astronomical Research Institute of Thailand (Public Organization),\\
260 Moo 4, T. Donkaew, A. Maerim, Chiangmai, Thailand}
\affiliation[g]{Institute of Cosmology and Gravitation, University of Portsmouth\\
Dennis Sciama Building, Burnaby Road, Portsmouth, PO1 3FX, United Kingdom}
\emailAdd{luhq@ihep.ac.cn,yupeng@sut.ac.th}
\abstract{
The Jiangmen Underground Neutrino Observatory (JUNO) is a 20 kton large liquid scintillator detector.JUNO's goals are the determination of neutrino mass ordering study and other neutrino physics research. The whole detector is 43.5 m in diameter and 44.0 m height. 18000 $20$ inches photomultiplier tubes(PMTs) used for photons detection in the central detector are sensitive and easily affected by the geomagnetic field. The PMT detection efficiency loss is about 60$\%$ under the geomagnetic field intensity ($\sim$0.5G). It significantly negatively impacts the detector performance, and a compensation system is required for geomagnetic field shielding. A system using 32 circular coils is chosen for shielding. The study shows that the residual magnetic field is less than 0.05 G in the Central Detector PMT region and can meet the experiment requirement. A prototype coil system with a 1.2 m dimension was built to validate the study and the design. The measured data of prototype and simulation results are consistent with each other, and geomagnetic field intensity is effectively reduced by coils, verifying the shielding coils system design for JUNO. 
}
\begin{document}
\maketitle
\section{Introduction}
\label{sec:intro}
Photomultiplier tubes(PMTs) are vacuum electronic devices that convert weak optical signals into electrical signals. These are used in optical measuring instruments and spectral analysis instruments. They played a vital role in high-energy physics experiments in the past decades. The Jiangmen Underground Neutrino Observatory (JUNO) is located at Kaiping, Jiangmen City, Guangdong Province, China~\cite{1,2,3}. The experiment uses a cylindrical detector with 43.5 m in diameter and 44.0 m in height. Inside the cylinder, the central detector (CD) is an acrylic sphere of 35.4 m in diameter filled with a 20 ktons liquid scintillator. There will be 18,000 20-inch Microchannel Plate Photomultiplier tubes (MCP-PMTs), and Hamamatsu PMTs \cite{4} installed in the central detector, combined with 26,000 3-inch PMTs to extend the dynamic range. Outside the CD, it will be filled with ultra-pure water, and 2,400 20-inch MCP-PMTs placed to serve as a water Cherenkov detector (Veto). The CD-PMTs and Veto-PMTs are installed in the range of 38.5 -- 41.1 m in diameter. The experiment needs to reach 3$\%$@1MeV energy resolution for physics requirement, which is the best energy resolution in liquid scintillator until now. To reach this goal, the PMT must have high detection efficiency and be in good working conditions.

When optical light is irradiated to the PMT photo-cathode, photo-electrons are excited into the vacuum of PMTs. Their trajectory will be deflected by Lorentz force due to the existing geomagnetic field. This effect will lead to apparent photo-electron collection efficiency loss in 20-inch PMTs \cite{5}. The collection efficiency loss depends strongly on the angle between the direction of the geomagnetic field and the moving direction of photo-electrons. The worst case is that the PMT orientation, defined as the vector from the PMT multiplier pole to the top of the PMT photocathode, is perpendicular to the direction of the geomagnetic field. When the PMT orientation is parallel to the geomagnetic field, it is the best case with no noticeable efficiency loss for PMTs.
Measurements have been performed to investigate the efficiency of PMTs in magnetic fields, where the PMT is placed in a coil system with controlled magnetic field intensity. The PMT photocathode is illuminated by the light from an LED diffuse ball, emitting different light intensities above the PMT.
\begin{figure}[htbp]
\centering 
\includegraphics[width=9cm]{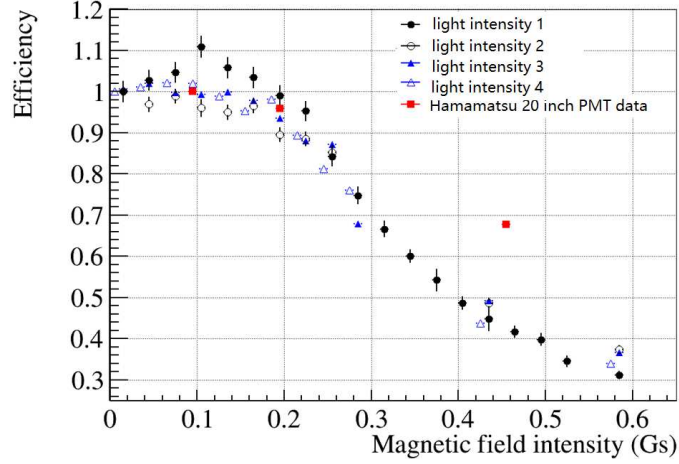}
\caption{\label{fig:1} The relationship between the magnetic field intensity and the detection efficiency of $20$-inch PMT.The magnetic field is at 90 degrees with respect to the PMT. The PMT detection efficiency is defined as 100$\%$ without a magnetic field. The red square is the data obtained from the Hamamatsu PMT datasheet, while the others are from MCP PMT measurements, where different symbols denote different LED intensities. The black hollow makers and solid markers are the measurements of a single photo-electron, the average observed photo-electron number 0.16, 0.38 respectively. While the blue hollow and solid triangles are for multi-photoelectron measurements with large signals of amplitude 40mV and 90 ~mV separately.
}
\end{figure}

Figure~\ref{fig:1} shows the PMT efficiency measured versus the residual magnetic field intensity, where the angle between the PMT orientation and the magnetic field direction is 90 degrees. The efficiency is the ratio of the observed photo-electron number by PMT between without and with a magnetic field at the same light intensity. The PMT detection efficiency is defined as 100$\%$ without a magnetic field. The red squares are the data obtained from the Hamamatsu PMT datasheet, while the others are the measurements of MCP-PMT at different light intensities. The PMT is operated at the gain of 10$^{7}$. The black hollow and solid marks are the measurements at low light intensity levels, with the average observed photo-electron numbers 0.16 and 0.38, respectively. In comparison, the blue hollow and solid triangles are multi-photoelectron with large signals of amplitude 40mV and 90 mV, respectively.

The figure shows that the efficiency loss due to magnetic field is similar for the MCP-PMT at different light intensities levels. The detection efficiency of the MCP-PMT is about 40$\%$ at 500 mG magnetic intensity. There is no apparent efficiency loss with lower intensity of less than 100 mG.

Other angles measurements found that the detection efficiency is minimum at 90 degrees, and the maximum efficiency is 1.02$\pm$0.03 at 0 degrees. 90 degrees result is used for a conservative evaluation of the geomagnetic influence. The experiment requires, based on these results that the field intensity in the CD region is less than 10$\%$ of the geomagnetic field (<50 ~mG).

To shield the magnetic field, we can apply compensation coils or high permeability materials such as mu-metal. If the PMT size is small (generally 2-inch, 3-inch, or 8-inch), the influence of the geomagnetic field can be reduced to a low level by mu-metal. For instance, the geomagnetic field's impact on the detection efficiency of the 8-inch PMT used in the Daya Bay experiment was controlled to less than 5$\%$ by mu-metal~\cite{6}.

For the mu-metal choice, one way is to use a mu-metal mesh to cover PMT individually, and another way is to choose a mu-metal sheet to cover the whole detector. The studies reveal that the mu-metal shielding's effectiveness is reduced for low-intensity fields in a vast dimensioned space~\cite{7,8}. A mu-metal mesh applied to each PMT can reduce the magnetic field intensity down to 150 mG, but it blocks some light for PMT because of mesh geometry~\cite{9,10}. Besides, the mu-metal is delicate and corrodes in pure water. Given that in JUNO, the PMTs will be placed in pure water, water pollution from the mu-metal will be a significant risk. The compensation coils, however, can be used in ample space. Consequently, the JUNO experiment plans to use compensation coils to compensate for the geomagnetic field penetrating the detector \cite{11}.

\indent The remainder of this paper is organized as follows. In Section~\ref{sec:design}, we focus on the design of the coils. In Section~\ref{sec:emfvar}, we show how the earth's magnetic field variation affects the design. In Section~\ref{sec:prototypesim}, we detail the design of the coils prototype. Section~\ref{sec:prototypeexp} is for the prototype setup and measurement. Conclusions are given in Section~\ref{sec:summary}.
\section{Compensation coil design for JUNO}
\label{sec:design}

\begin{figure}[htbp]
\begin{center}
\includegraphics[width=10cm]{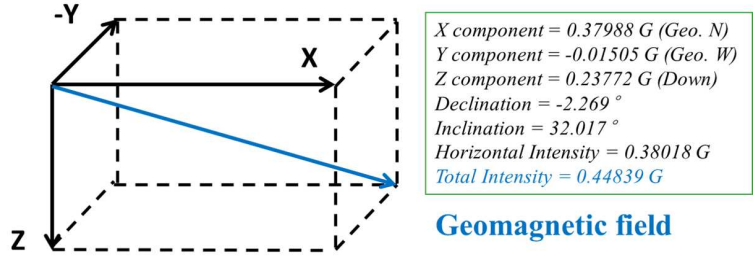}
\vspace{0cm}
\caption{\label{fig:2} The geomagnetic field intensities in Jiangmen.}
\end{center}
\end{figure}

The earth's geomagnetic field is everywhere, with different intensities at different locations. Detailed information can be found in the International Geomagnetic Reference Field (IGRF), edited by the International Association of Geomagnetism and Aeronomy (IAGA)~\cite{12}. The JUNO experiment is near Jiangmen city, China, with a latitude of 22.127 degrees north and 112.517 degrees east. The IGRF data of the earth's magnetic field at the JUNO site are shown in Figure ~\ref{fig:2}. The geomagnetic field intensity in Jiangmen is 380.18 mG in the Geographic North Pole horizontal direction and 237.72 mG in the vertical direction. The angle between the magnetic field direction and the horizontal direction is 32.017 degrees. Underground measurements will be carried out in the future when the civil construction of the experiment hall and the detector pool is completed to measure the exact magnetic field on-site rather than in the general area as given by the IGRF.

Until now, except for the Super-Kamiokande experiment~\cite{13}, there is no other experiment using active geomagnetic shielding for 20-inch PMTs in such an ample space. The Super-Kamiokande detector is a water Cherenkov detector, where there was enough space to install coils outside the detector. The requirement for magnetic shielding is not as strict as the JUNO requirement. For JUNO, outside the acrylic sphere is the veto detector without extra space for the coil installation. The coils must be installed inside the pool and generate a uniform magnetic field in a vast region. To simplify the installation, we should use as few coils as possible.
\begin{figure}[htbp]
\centering 
\includegraphics[width=9cm]{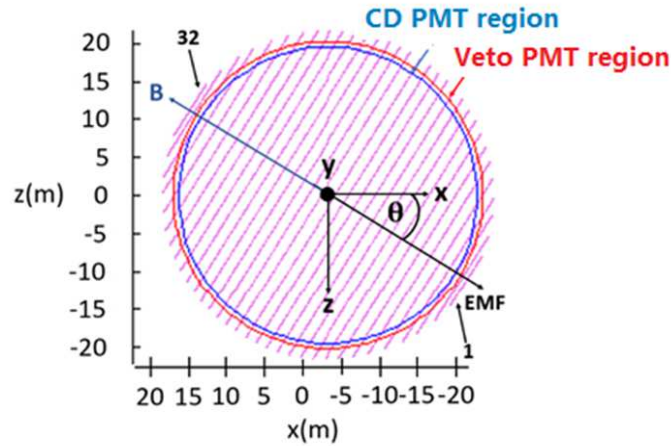}
\caption{\label{fig:3} The schematic diagram of JUNO active geomagnetic shielding coils (sectional view of the coils). A total of 16 pairs of coils will be installed.}
\end{figure}
\indent

We have carried out comprehensive studies and optimizations on different coil layouts. Finally, one set of coils was chosen for the detector, which can simplify the installation and control the cost, as shown in Figure~\ref{fig:3}. The shielding coils are arranged in parallel rings placed on the surface of a sphere of 43.5 m diameter. The goal is to reduce the magnetic field to less than 10$\%$ of the geomagnetic field intensity in the CD-PMT region between 38.5 m and 39.5m in diameter and below 20$\%$ in the Veto-PMT area between 40.6 m and 41.1 m in diameter.
The magnetic induction generated by a Direct Current coil is evaluated with Biot-Savart's law. In the numerical calculation, a circular coil shown in Figure~\ref{fig:3} is cut into 1,000 sections, each treated as a small straight line, and the line integration is performed by applying the Gauss-Legendre formula~\cite{14},
\begin{equation}
\label{eq:bs}
\bm{B} = \frac{\mu_{0} I}{4 \pi} ~\sum_{q=1}^{n} \frac{\bm{\alpha}}{r_{pq}^{3}}
\end{equation}
with
\begin{equation}
\label{eq:bscom}
\begin{aligned}
\bm{\alpha} &= \alpha_x \hat{i} + \alpha_y \hat{j} + \alpha_z \hat{k}, \\
\alpha_x &= \Delta z_{pq}~w(y_q) - \Delta y_{pq}~w(z_q), \\ \alpha_y &= \Delta x_{pq}~w(z_q) - \Delta z_{pq}~w(x_q), \\ \alpha_z &= \Delta y_{pq}~w(x_q) - \Delta x_{pq}~w(y_q), \\
r_{pq} &= \sqrt{\Delta x_{pq}^2 + \Delta y_{pq}^2 + \Delta z_{pq}^2},
\end{aligned}
\end{equation}
where $\Delta x_{pq} = x_p - x_q,~ \Delta y_{pq} = y_p - y_q,$ and $\Delta z_{pq} = z_p - z_q$ with ($x_{q}$, $y_{q}$, $z_{q}$) being the coordinates of a section of the coil and ($x_{p}$, $y_{p}$, $z_{p}$) being the coordinates at which the generated magnetic field is evaluated, $I$ is the current of the coil, and $w(x_q)$, $w(y_q)$ and $w(z_q)$ are the weights given in the Cartesian coordinates in Gauss-Legendre formula. The minimum number of coils, the intervals between two neighboring coils, and each coil current are optimized to meet the experiment requirement. The residual intensity ($RI$) is defined as
\begin{align}
RI = \frac{\sqrt{\left(B_x - EMF_{x}\right)^2 + \left(B_y - EMF_{y}\right)^2 + \left(B_z - EMF_{z}\right)^2}}{EMF} \times 100 \%,
\end{align}
where $EMF$, $EMF_{x}$, $EMF_{y}$, and $EMF_{z}$ are the absolute value, the northern, the eastern, and the geomagnetic field's vertical components, respectively, and the magnetic inductions in the Cartesian coordinates from the coils are given by $B_x$,$B_y$,$B_z$. \\
\indent Various layouts of coils have been studied, such as $14$, $15$, $16$, and $17$ pairs of coils with equal and unequal spacings. The currents have been optimized using the linear constrained least-squares method, the so-called bounded-variable least squares (BVLS), using the lsq$\_$linear function Python-based system of open-source software named from Scipy \cite{15}. In this work,the coil currents were optimized to minimize $\lvert Ax-b\rvert$,
\begin{equation}
A =
\begin{pmatrix}
Bx_{11} & Bx_{12} & \cdots & Bx_{1n} \\
By_{11} & By_{12} & \cdots & By_{1n} \\
Bz_{11} & Bz_{12} & \cdots & Bz_{1n} \\
Bx_{21} & Bx_{22} & \cdots & Bx_{2n} \\
By_{21} & By_{22} & \cdots & By_{2n} \\
Bz_{21} & Bz_{22} & \cdots & Bz_{2n} \\
\vdots & \vdots & \ddots & \vdots \\
Bx_{m1} & Bx_{m2} & \cdots & Bx_{mn} \\
By_{m1} & By_{m2} & \cdots & By_{mn} \\
Bz_{m1} & Bz_{m2} & \cdots & Bz_{mn} \\
\end{pmatrix},\;\;
~x =
\begin{pmatrix}
I_{1} \\
I_{2} \\
I_{3} \\
I_{4} \\
I_{5} \\
I_{6} \\
\vdots \\
I_{n-2} \\
I_{n-1} \\
I_{n} \\
\end{pmatrix},\;\;
~b =
\begin{pmatrix}
EMFx_{1} \\
EMFy_{1} \\
EMFz_{1} \\
EMFx_{2} \\
EMFy_{2} \\
EMFz_{2} \\
\vdots \\
EMFx_{m} \\
EMFy_{m} \\
EMFz_{m} \\
\end{pmatrix},
\end{equation}
where $Bx_{ij}$, $By_{ij}$, and $Bz_{ij}$ are the magnetic inductions in the Cartesian coordinates of the $i^{th}$ field point computed from Equation~\eqref{eq:bs} produced by the $j^{th}$ circular coil, $EMFx_{i}$, $EMFy_{i}$, and $EMFz_{i}$ are the geomagnetic field in the $i^{th}$ field point in the $x$, $y$ and $z$ directions,
and $I_{j}$ is the current of the $j^{th}$ circular coil. In the optimization, $32,000$ points are chosen in the CD-PMT and Veto-PMT regions.

\indent Finally, a set of 16 pairs of circular coils with almost equal spacings can efficiently compensate for the geomagnetic field. The residual magnetic field intensity is less than $10\%$ of the geomagnetic field intensity in the CD-PMT region and below $20\%$ in the Veto-PMT region. The dimensions of the 16 pairs and the optimized currents are shown in Table~\ref{tab:i}, where $R_c$ and $Z_c$ are the coordinates of the coil centers in the radial and axial directions, respectively, and $\Delta Z_{i,i+1}$ are the intervals between two neighboring coils.
\begin{table}[h]
\caption{\label{tab:i} 16 pairs of geomagnetic field compensation coil parameters for JUNO.}
\vspace{5 mm}
\centering
\footnotesize
\begin{tabular}{|c c c c c | c c c c c|}
\hline
Coil Num.&$R_c$/m&$Z_c$/m&$\Delta Z_{i,i+1}$/m&I/A\\
\hline
1 & 3.90 & $\pm$21.30 & - & 26.36 \\
2 & 6.57 & $\pm$20.63 & 0.67 & 52.72 \\
3 & 9.25 & $\pm$19.58 & 1.05 & 52.72 \\
4 & 11.12 & $\pm$18.58 & 1.00 & 65.90 \\
5 & 13.28 & $\pm$17.10 & 1.48 & 79.08 \\
6 & 14.99 & $\pm$15.62 & 1.48 & 79.08 \\
7 & 16.39 & $\pm$14.14 & 1.48 & 79.08 \\
8 & 17.57 & $\pm$12.66 & 1.48 & 79.08 \\
9 & 18.54 & $\pm$11.18 & 1.48 & 79.08 \\
10 & 19.36 & $\pm$9.70 & 1.48 & 79.08 \\
11 & 20.03 & $\pm$8.22 & 1.48 & 79.08 \\
12 & 20.57 & $\pm$6.74 & 1.48 & 79.08 \\
13 & 21.01 & $\pm$5.26 & 1.48 & 79.08 \\
14 & 21.32 & $\pm$3.78 & 1.48 & 79.08 \\
15 & 21.53 & $\pm$2.30 & 1.48 & 79.08 \\
16 & 21.63 & $\pm$0.82 & 1.48 & 79.08 \\
\hline
\end{tabular}
\end{table}

\begin{center}
\begin{figure}[!ht]
\begin{minipage}{14pc}
\centering
\includegraphics[width=18pc]{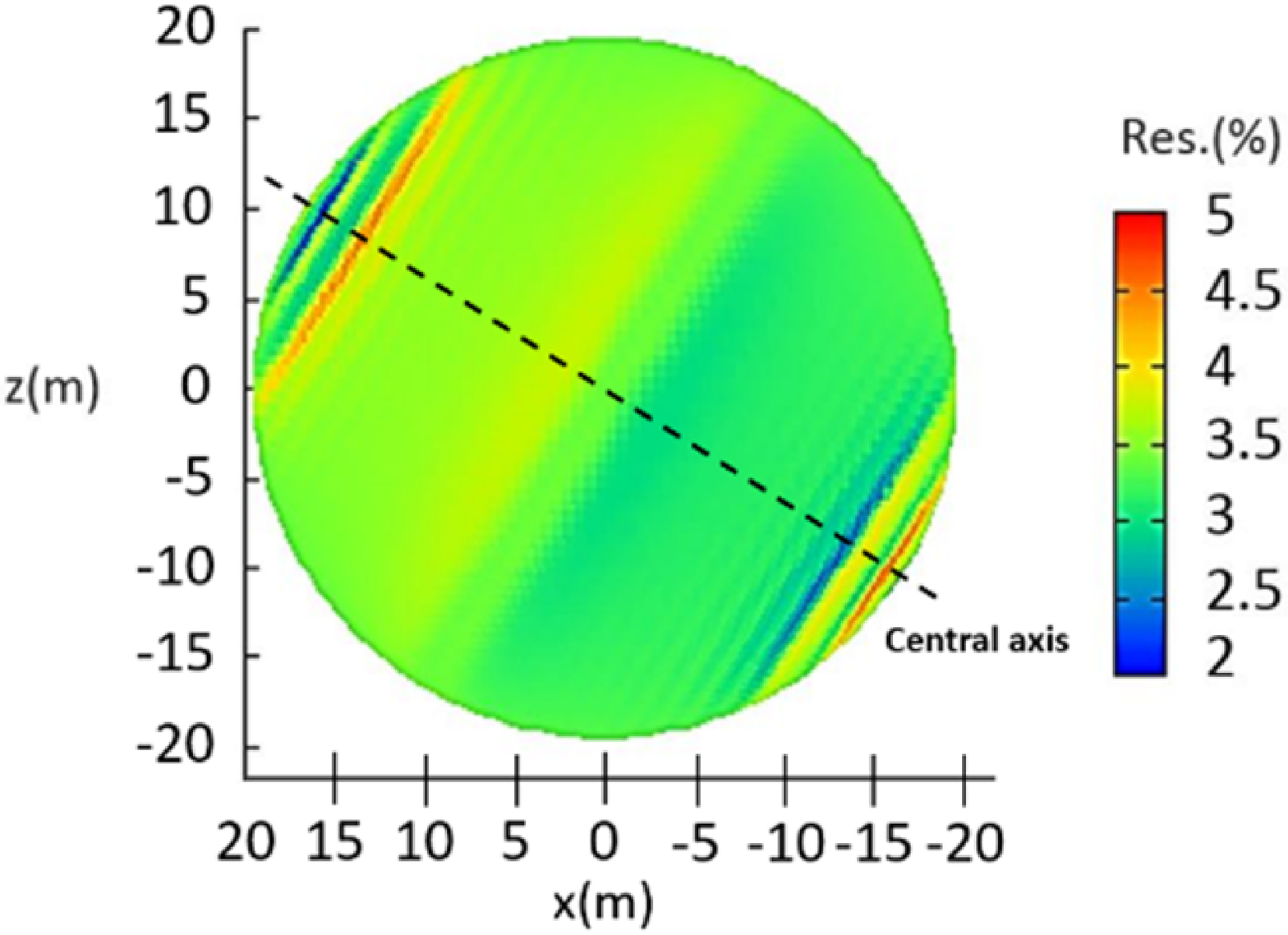}
\caption{\label{fig:4} $RI$ on the spherical surface (side view) in the CD-PMT region.}
\end{minipage} \hspace{6pc}%
\begin{minipage}{14pc}
\centering
\includegraphics[width=18pc]{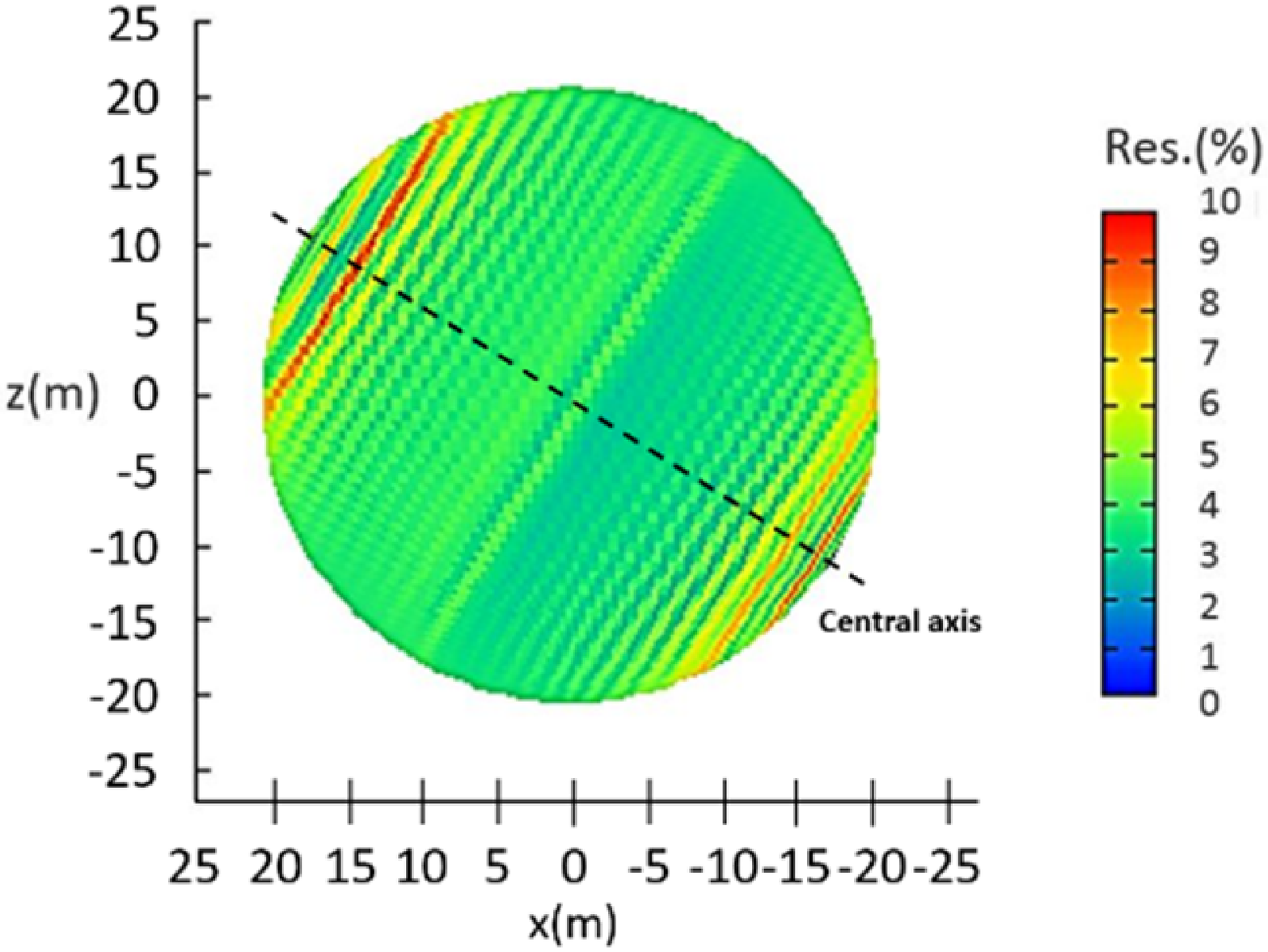}
\caption{\label{fig:5} $RI$ on the spherical surface (side view) in the Veto-PMT region.}
\end{minipage}
\end{figure}
\end{center}

\begin{center}
\begin{figure}[!ht]
\begin{minipage}{14pc}
\centering
\includegraphics[width=18pc]{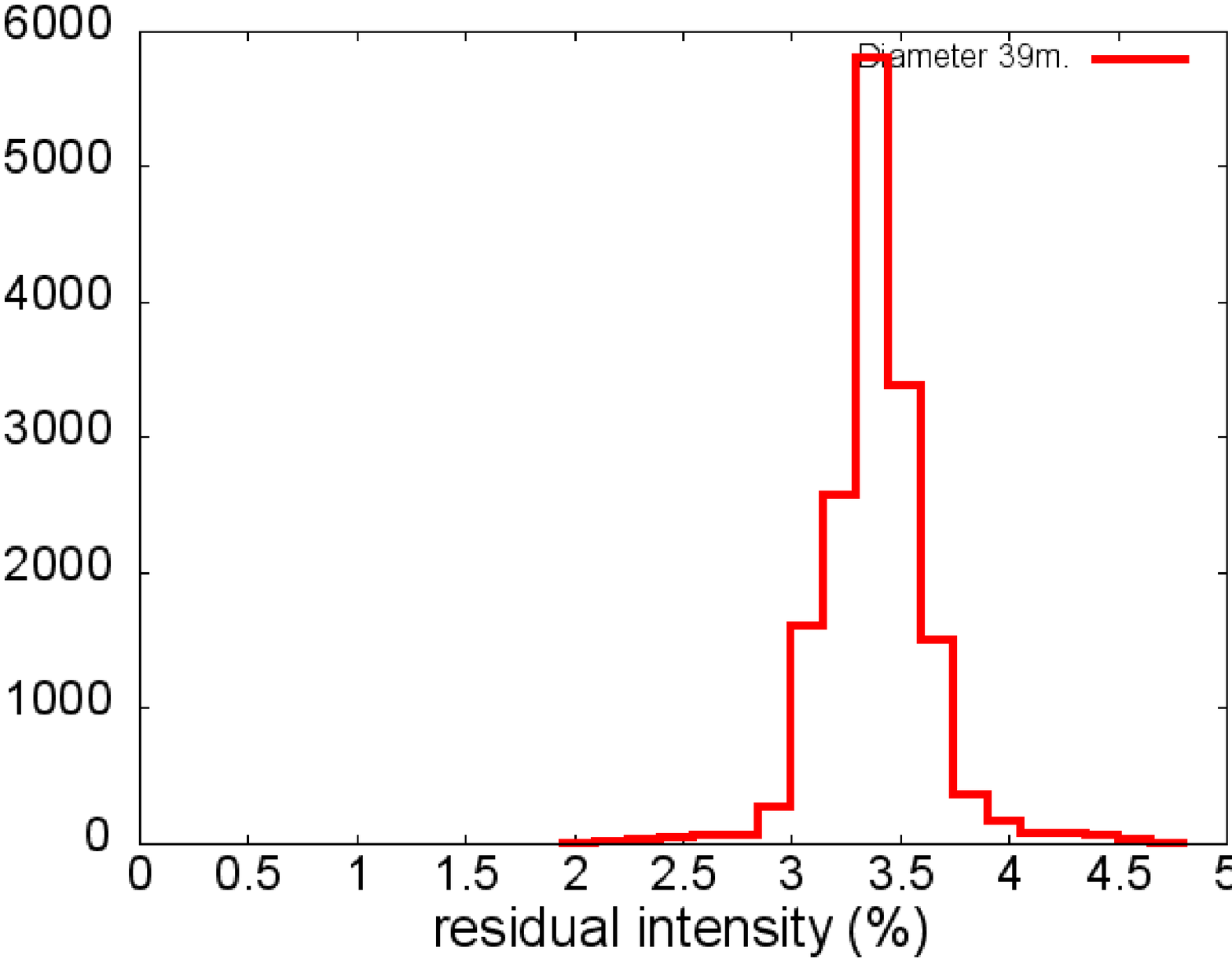}
\caption{\label{fig:6} $RI$ distribution on the spherical surface of diameter 39 m (CD-PMT region).}
\end{minipage} \hspace{6pc}%
\begin{minipage}{14pc}
\centering
\includegraphics[width=18pc]{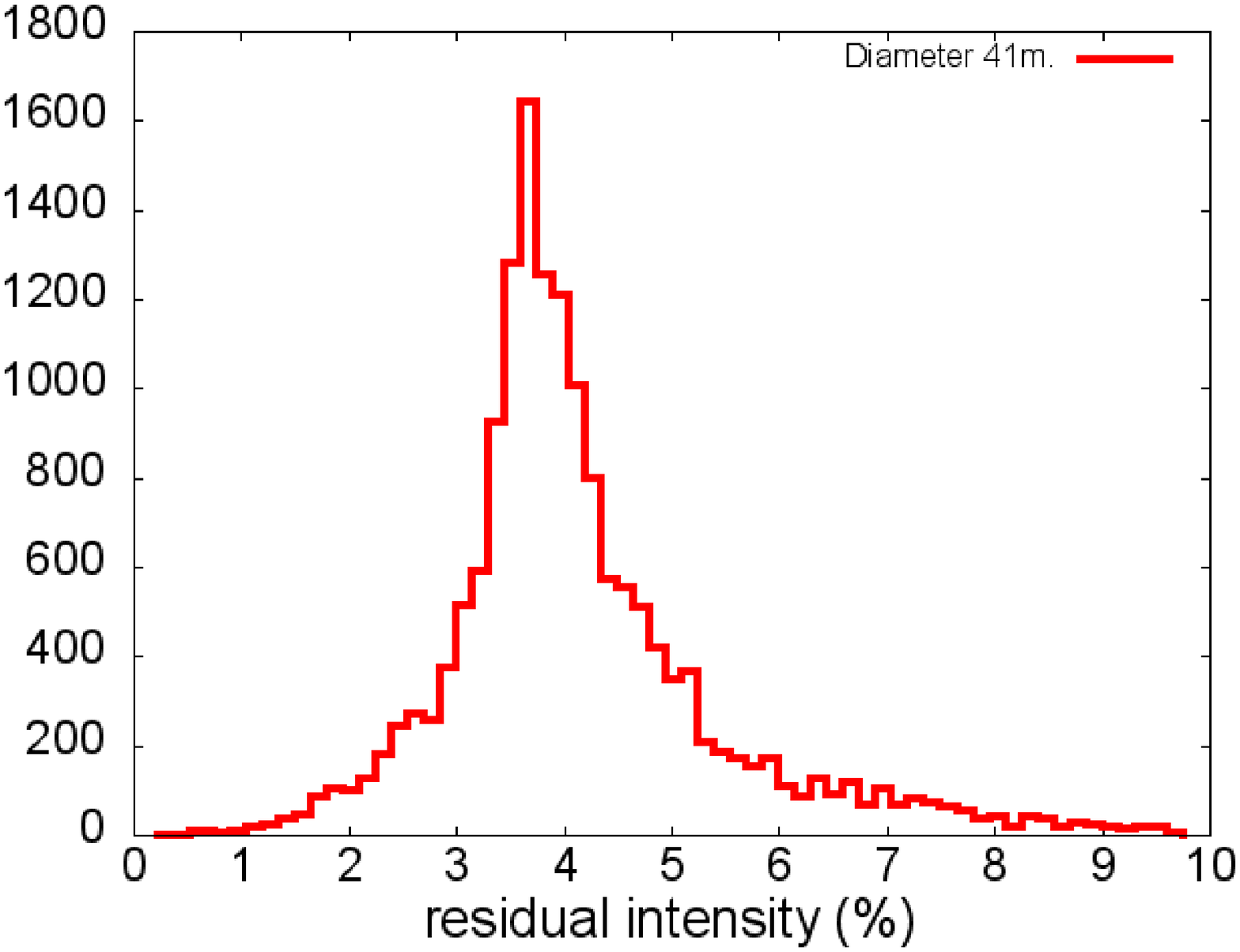}
\caption{\label{fig:7} $RI$ distribution on the spherical surface of diameter 41 m (Veto-PMT region).}
\end{minipage}
\end{figure}
\end{center}

The 16 pairs of coils generate a magnetic field with almost the same intensity as the geomagnetic field but opposite direction. With these coils, the geomagnetic field along the central axis can be well shielded. However, there is still some residual magnetic field at positions away from the axis, and the $x$, $y$, and $z$ components of the residual magnetic field vary with positions. Nevertheless, the residual magnetic field intensity is minimal. Therefore, the total residual intensities in the CD-PMT and Veto-PMT regions are used instead of the residual magnetic field components as a conservative estimation, shown in Figure~\ref{fig:4} and Figure~\ref{fig:5}, respectively. The $RI$ distribution on the spherical surface of 39~m and 41~m diameter regions from the Figure~\ref{fig:4} and the Figure~\ref{fig:5} are converted to histogram plots in Figure~\ref{fig:6} and Figure~\ref{fig:7} \cite{16}. The maximum $RI$ is $\pm$4.66$\%$ of the geomagnetic field on the 39.0-m diameter surface and $\pm$9.68$\%$ of the geomagnetic field on the 41~m diameter surface.
\section{Geomagnetic field variations}
\label{sec:emfvar}

\begin{figure}[!h]
\begin{center}
\includegraphics[width=9cm]{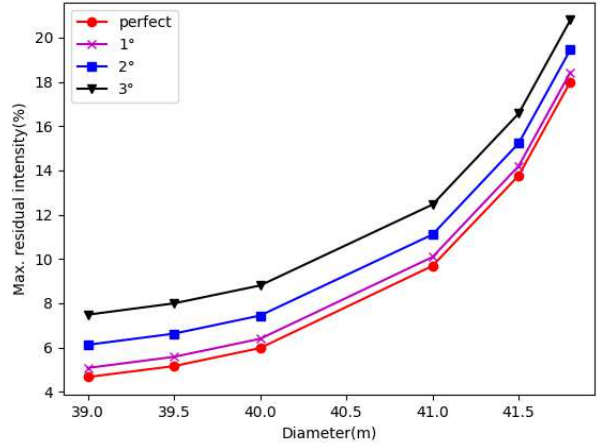}
\end{center}
\caption{\label{fig:8}Maximum residual intensities on the spheres of diameter $38.5$, $39.0$, $39.5$, $40.0$, $41.0$ and $41.5$ m when the inclination angles of the EMF are $1^{\circ}$, $2^{\circ}$ and $3^{\circ}$.}
\end{figure}

The geomagnetic field varies year by year and the IGRF data in the last 20 years (1997-2017) show that the magnetic declination and magnetic inclination have changed respectively ${-0.0591^{\circ}}$ and $0.1340^{\circ}$ per year in Jiangmen~\cite{17}, which means the geomagnetic field in Jiangmen will change approximately 2.68 degrees in the next 20 years. The field changes with three inclination angles of $1^{\circ}$, $2^{\circ}$, and $3^{\circ}$, normal to the $y$-axis in the counterclockwise direction, as listed in Table~\ref{tabii}. Therefore, it is necessary to consider the coils' shielding effect in which the field varies a few degrees. The residual intensities are estimated with the inclination angle of the field varying $1^{\circ}$, $2^{\circ}$, and $3^{\circ}$. The maximum residual intensities on different spherical surfaces are shown in Figure~\ref{fig:8}. Therefore, the coils can still shield the geomagnetic field sufficiently for both the CD-PMT and Veto-PMT for the next 20 years.
\begin{table}[h]
\caption{ \label{tabii} Geomagnetic field variation with three inclination angles (units: mG).}
\vspace{5 mm}
\centering
\begin{tabular}{|c c c c|}
\hline
Direction & $1^{\circ}$ & $2^{\circ}$ & $3^{\circ}$ \\
\hline
North ($x$) & 383.97 & 387.94 & 391.80 \\
West ($y$) & 15.05 & 15.05 & 15.05 \\
Vertical ($z$) & 231.05 & 224.32 & 217.51 \\
\hline
\end{tabular}
\end{table}

Some further improvements could be applied to reduce the geomagnetic field variation effect. One way is to adjust the coil's currents. The direction of the shielding magnetic field cannot be changed after the coils are installed. One may optimize the coil currents to minimize the residual intensity when the local magnetic field direction deviates. However, the allowed current change of the coils is minimal (<0.2 $\%$), and the reduction effect is limited. Another way is to set a compensation angle when the coils are installed. The first ten years are the critical period for the JUNO experiment, and we will introduce an angle compensation for five years. The coil axis declination and inclination angles will be added by$ -(-0.0591 \times 5$) $=0.3$ and $-(0.1340 \times 5$) $=-0.7$ degrees, respectively, for the coil installation. After the angle compensation, the angle difference between the geomagnetic field direction and the coil axis direction is lower than $1^{\circ}$ during the first ten years.
\section{Design of prototype coil system}
\label{sec:prototypesim}

A small coil prototype system was built to verify the design of the compensation coils for JUNO. In order to increase the adaptability and flexibility of magnetic field control, two sets of prototype shielding coils are installed to shield the geomagnetic field's horizontal and vertical components. The outer diameter and height of the prototype coil system are 1.2 m. We have employed the same program \cite{14,18} to optimize the shielding. The purpose of the optimization is to minimize the wire consumption and the peak-to-peak inhomogeneity of the magnetic field intensity to less than 5$\%$ in the central region within 0.5 meters. We also have carried out a PMT performance test under different magnetic intensities by installing a PMT in the coils simultaneously.
\subsection{Coils of vertical magnetic field component $B_1$}

The parameters of the coils for shielding the vertical magnetic field component $B_1$ are obtained as shown in Table~\ref{tab:iii}. The set consists of 7 coils distributed symmetrically along the vertical direction. The first and last columns of the table are the coil label and the number of turns of wire used in each coil. $R_c$ and $Y_c$ are the coils' radius and coordinates in the axial direction, respectively.
\begin{table}[h]
\caption{\label{tab:iii} Coils parameters for vertical magnetic field component $B_1$.}
\centering
\footnotesize
\begin{tabular}{|c c c c | c c c c|}
\hline
Coil Num.&$R_c$/m&$Y_c$/m&Turns\\
\hline
1 & 0.380 & -0.600 & 5 \\
2 & 0.600 & -0.350 & 8 \\
3 & 0.600 & -0.117 & 3 \\
4 & 0.600 & 0.000 & 2 \\
5 & 0.600 & 0.117 & 3 \\
6 & 0.600 & 0.350 & 8 \\
7 & 0.380 & 0.600 & 5 \\
\hline
\end{tabular}
\end{table}
\subsection{Coils of horizontal magnetic field component $B_2$}
The coils for horizontal component shielding are installed inside the vertical shielding coils. The optimization algorithm for the $B_2$ coils is the same as for the $B_1$ coils. The obtained coil parameters are shown in Table~\ref{tab:iv} in the same manner as Table~\ref{tab:iii}. The horizontal shielding set consists of 6 pairs of coils distributed symmetrically along the horizontal direction.
\begin{table}[h]
\caption{\label{tab:iv} Coil parameters for horizontal magnetic field component $B_2$.}
\centering
\footnotesize
\begin{tabular}{|c c c c | c c c c|}
\hline
Coil Num.&$R_c$/m&$X_c$/m&Turns\\
\hline
1 & 0.570 & $\pm$0.053 & 5 \\
2 & 0.548 & $\pm$0.163 & 5 \\
3 & 0.505 & $\pm$0.271 & 3 \\
4 & 0.434 & $\pm$0.366 & 2 \\
5 & 0.342 & $\pm$0.456 & 3 \\
6 & 0.204 & $\pm$0.540 & 8 \\
\hline
\end{tabular}
\end{table}
\subsection{Simulation results}

\begin{figure}[htbp]
\begin{center}
\includegraphics[width=8cm]{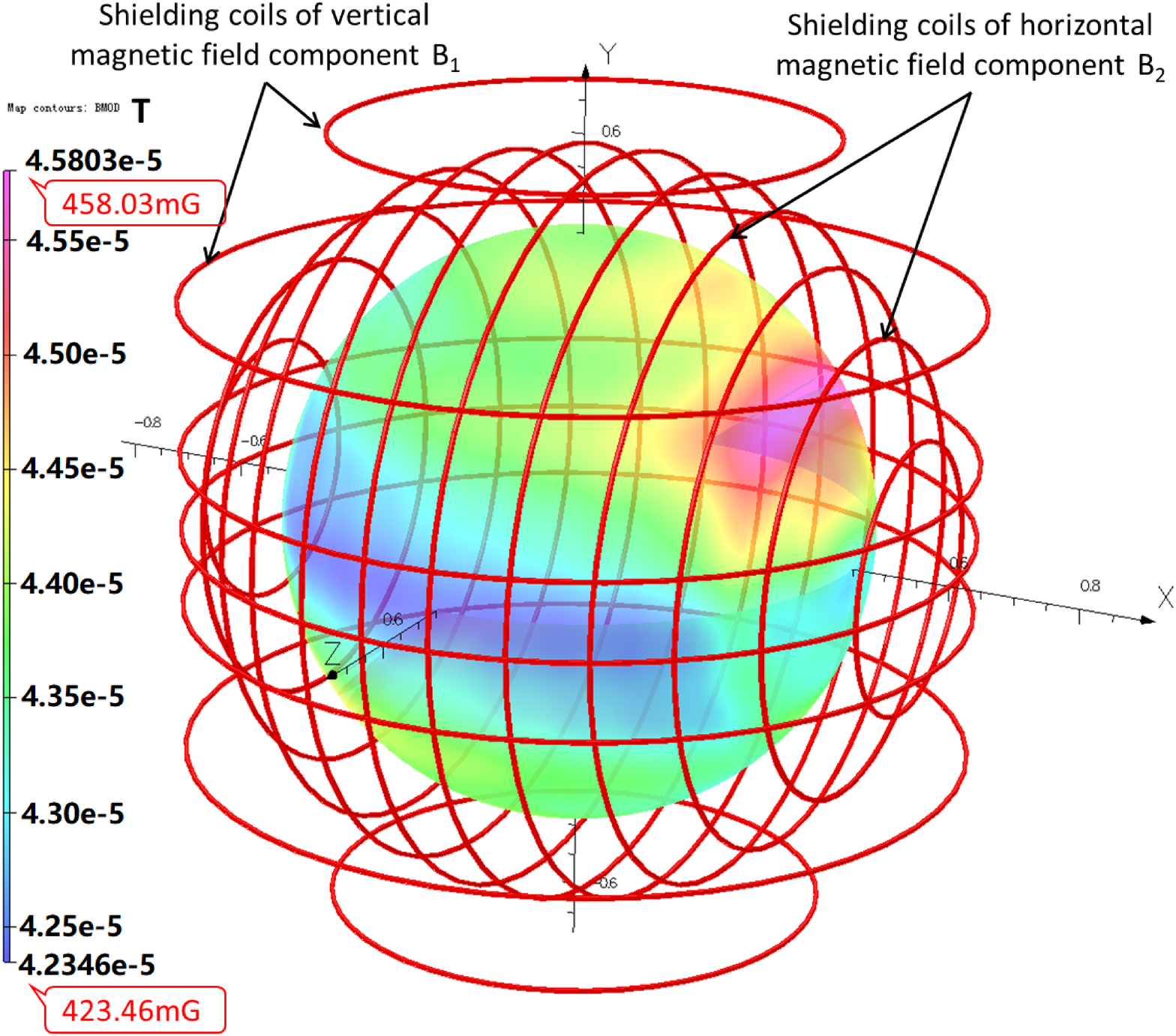}
\vspace{0cm}
\caption{\label{fig:9} Distribution of magnetic induction intensity on a sphere surface with a diameter of 1000 mm (Axis (X,Y,Z) given in mm and color scale for field intensity given in T).}
\end{center}
\end{figure}

According to the measurements, the laboratory's total geomagnetic field intensity is 433~mG, with the horizontal and vertical components being 308 mG and 305 mG, respectively. The operating currents of the horizontal and vertical shield coils are determined to be respectively 0.795 A and 1.430 A. Shown in Figure~\ref{fig:9} is the distribution of the magnetic field intensity on the sphere surface with a diameter of 1.0~m inside the shielding coils. Simulation results show that the magnetic field intensity at the center of the coil is 433 mG. The peak-to-peak magnetic intensity deviations are less than 8$\%$ on the 1.0 m diameter sphere surface and less than 1$\%$ in the central region within 0.5 m.
\section{Prototype coil system}
\label{sec:prototypeexp}
\begin{figure}[htbp]
\begin{center}
\includegraphics[width=10cm]{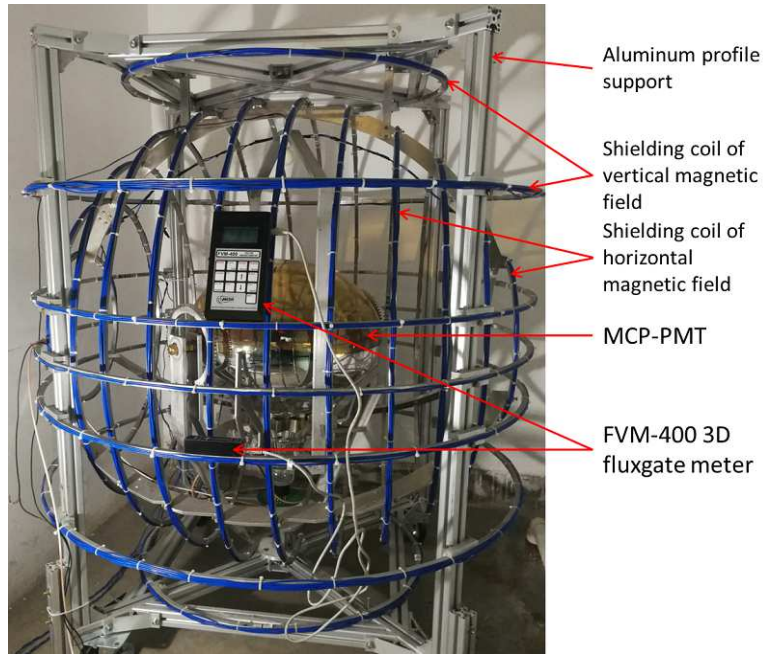}
\vspace{0cm}
\caption{\label{fig:10} The geomagnetic shielding coil system.}
\end{center}
\end{figure}

The support structure of the coil system is made of aluminum alloy. Non-magnetic brass bolts and other non-magnetic materials for fixing the coils are used to minimize magnetic material's influence on the geomagnetic field distribution. In addition, the $20$-inch PMT rotatable measuring device is integrated into the shielding coil system. The coil system and the $20$-inch PMT measurement device are shown in Figure~\ref{fig:10}.
\subsection{Data measurement}
A Henki HT2320 three-channel programmable DC power supply was used to supply power to the shielding coils, and a FVM-400 handheld vector 3D fluxgate meter made by Macintyre Electronic Design Associates was used for measuring the geomagnetic field intensity.
\begin{figure}[htbp]
\begin{center}
\includegraphics[width=10cm]{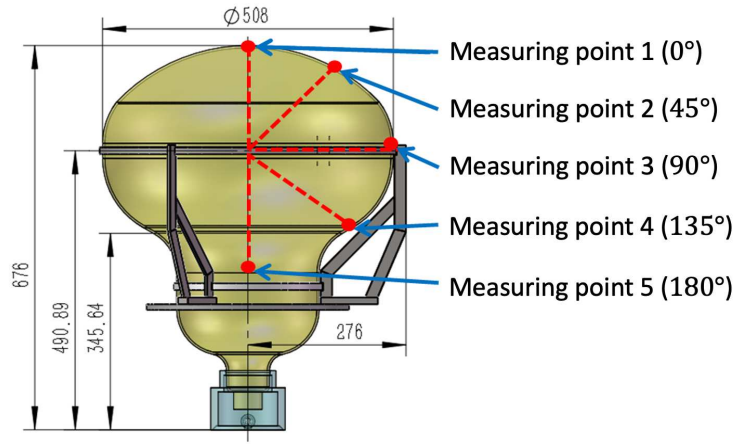}
\vspace{0cm}
\caption{\label{fig:11} Distribution of magnetic field measurement points on 20-inch PMT spherical surface (unit: mm).}
\end{center}
\end{figure}
As shown in Figure~\ref{fig:10}, the 20-inch PMT is located at the center of the shielding coils within 0.5 meters. The magnet intensity was measured at 26 positions on the surface of the PMT under different conditions, as shown in Fig.~\ref{fig:11} Table~\ref{tab:v}. These measured data are listed in Columns 3 and 5 in Table~\ref{tab:v}, while the program calculation's simulation values are shown in Column 2. The measured data include both before and after switching on the coils. As shown in Column 5 of Table~\ref{tab:v}, when the shielding coils are energized, the measured residual magnetic field intensity is lower than 24 mG at each measuring point, which means that over 95$\%$ of the geomagnetic field has been compensated.

\begin{table}[h]
\caption{\label{tab:v} Results of magnetic field intensity at different positions under different conditions.}
\centering
\footnotesize
\begin{tabular}{|c c c c c c | c c c c c c|}
\hline
Measurement & Simulated & Measured & Theoretical & Measured & Absolute \\
point & magnetic field /mG & geomagnetic & absolute & residual field & deviation$\ast$/ \\
(R,$\Theta$,$\Phi$)/mm & & field/mG & deviation/mG & /mG & mG \\
\hline
200,0,0 & 433.16 & 438.2 & 5.04 & 4.4 & 0.64 \\
220,45,0 & 433.12 & 433.3 & 0.18 & 6.1 & 5.92 \\
220,45,45 & 433.11 & 434.7 & 1.59 & 7.2 & 5.61 \\
220,45,90 & 433.10 & 435.2 & 2.10 & 4.8 & 2.70 \\
220,45,135 & 433.07 & 436.8 & 3.73 & 6.6 & 2.87 \\
220,45,180 & 433.06 & 445 & 11.94 & 9.8 & 2.14 \\
220,45,225 & 433.07 & 443.2 & 10.13 & 8.6 & 1.53 \\
220,45,270 & 433.10 & 435.6 & 2.50 & 5.2 & 2.70 \\
220,45,315 & 433.11 & 433.9 & 0.79 & 4.6 & 3.81 \\
265,90,0 & 432.96 & 423.7 & 9.26 & 16.2 & 6.94 \\
265,90,45 & 432.98 & 422.2 & 10.78 & 13.4 & 2.62 \\
265,90,90 & 432.96 & 424 & 8.96 & 12.4 & 3.44 \\
265,90,135 & 432.98 & 441 & 8.02 & 13.5 & 5.48 \\
265,90,180 & 432.96 & 446.4 & 13.44 & 13.4 & 0.04 \\
265,90,225 & 432.98 & 445.4 & 12.42 & 18 & 5.58 \\
265,90,270 & 432.96 & 430.2 & 2.76 & 13.7 & 10.94 \\
265,90,315 & 432.98 & 427.3 & 5.68 & 8.8 & 3.12 \\
220,135,0 & 433.00 & 422.3 & 10.70 & 19.6 & 8.90 \\
220,135,45 & 433.01 & 418.4 & 14.61 & 23.7 & 9.09 \\
220,135,90 & 433.02 & 432.1 & 0.92 & 12.6 & 11.68 \\
220,135,135 & 433.04 & 430.6 & 2.44 & 14.5 & 12.06 \\
220,135,180 & 433.06 & 446.8 & 13.74 & 16.3 & 2.56 \\
220,135,225 & 433.04 & 439.5 & 6.46 & 15.7 & 9.24 \\
220,135,270 & 433.02 & 424.3 & 8.72 & 23.9 & 15.18 \\
220,135,315 & 433.01 & 429.3 & 3.71 & 11 & 7.29 \\
200,180,0 & 433.16 & 436.5 & 3.34 & 17.8 & 14.46 \\
\hline
\end{tabular}
\\
{\footnotesize $^{\ast}$These data are the deviation between the calculated residual magnetic field and the measured residual magnetic field.}
\end{table}
\subsection{Discussions}

Column 2 in Table~\ref{tab:v} shows that the peak-to-peak magnetic intensity deviation derived from calculation results is less than 1$\%$ in the central region of the coils. The direction of the magnetic field is almost unique. However, the non-uniformity of measurements is larger than the ideal theoretical results, as shown in Column 3. Several factors need to be taken into account in the actual case.

One factor is that the magnetic materials such as steel bars inside the building near the experimental lab influence the geomagnetic field distribution inside the shielding coils. The geomagnetic field uniformity and direction are slightly altered, and thus the magnetic field generated by the shielding coil is not entirely parallel to the geomagnetic field. Column 3 in Table~\ref{tab:v} shows that the geomagnetic field deviation is up to 3$\%$. The coil installation positioning error is about 5 mm. The simulation results show that when the coil position and deformation errors are $\pm$5 mm, the maximum magnetic field uniformity deviation at the 20-inch PMT position is about 2$\%$. The accuracy of measurements is $\pm$0.5$\%$ for FVM-400 meters. These factors will induce about a 4$\%$ difference between the measured data and simulation results. If the simulation includes these actual cases, the absolute deviations for these positions will be about 5$\%$. As shown in Column 6 in Table~\ref{tab:v}, the deviations between the calculated and measured residual magnetic fields are less than 5$\%$. The data and theoretical simulation results are consistent with each other.

For the detector of JUNO in the future, the positioning accuracy of the coils is required to be less than 5 cm, and the error induced is less than 1$\%$. The error of the instruments of geomagnetic measurement is 0.5$\% - 1\%$. The uniformity of the geomagnetic field at the detector location will depend on future field measurements.
\section{Conclusions}
\label{sec:summary}

The geomagnetic field reduces the detection efficiency of large dimension PMTs significantly. After investigating different shielding approaches, JUNO chooses compensation coils to shield the geomagnetic field. After optimization studies on various layouts of compensation coils, a system of one set of 32 circular shielding coils will be used in the detector. The residual magnetic field intensity is less than 10$\%$ of the geomagnetic field intensity in the CD-PMT region and below 20$\%$ in the Veto-PMT region with the shielding coils.

\indent A prototype of the coil system was built, and the measured data and the simulation results are consistent. In the central region, the residual intensity is less than 5$\%$ of the geomagnetic field intensity, which shows that the active magnetic shielding system has a good shielding effect on the geomagnetic field. The study verifies the reliability and feasibility of the compensation coil system design for the JUNO experiment. Furthermore, this study could provide practical guidance for the design of future large-scale neutrino detectors for PMT magnetic field shielding.

The geomagnetic shielding coils system design was presented in this paper. 
The final positioning of coils will be done once on-site measurements of the field are performed. The entire installation process will be carried out together with the other parts of the detector, from top to bottom.
The JUNO detector installation will be completed in 2022, and the detector operation will start in 2023. 

\acknowledgments
The article is supported by the National Natural Science Foundation of China (Grant Nos. 11705219), the Key Intergovernmental Specialities in International Scientific and Technological Innovation Cooperation, MOST (No.2017YFE0132500), Suranaree University of Technology (SUT), IRD Full-time Master Researcher (contract no. Full-time 61/05/2562). The computing resource has been provided by the Center for Computer Services at SUT. CUniverse research promotion project of Chulalongkorn University (grant CUAASC), Rachadapisek Sompot Fund, Chulalongkorn University. The authors would like to acknowledge Thai-JUNO Consortium and IHEP EMF Shielding Group for instruction and comments.


\begin{thebibliography}{99}
\bibitem{1}
P. Devore et al., \emph{Light-weight flexible magnetic shields for large-aperture photomultiplier tubes}, \emph{Nuclear Instruments and Methods in Physics Research. Section A, Accelerators, Spectrometers, Detectors and Associated Equipment}, {\bf 737} (2014) pg. 222 $-$ 228.
\bibitem{2}
F. An et al., \emph{Neutrino physics with {JUNO}}, \emph{Journal of Physics G: Nuclear and Particle Physics} {\bf 43} (2016) pg. 030401.
\bibitem{3}
J. Cao, \emph{Daya Bay and Jiangmen Underground Neutrino Observatory (JUNO) neutrino experiments}, \emph{SCIENTIA SINICA Physica, Mechanica \& Astronomica} {\bf 44} (2014) pg. 1025 $-$ 1040.
\bibitem{4}
Y.-F. Li et al., \emph{Unambiguous determination of the neutrino mass hierarchy using reactor neutrinos}, \emph{Physical Review D}, {\bf 88} (2013) pg. 013008.
\bibitem{5}
Y. Wang et al., \emph{A new design of large area MCP-PMT for the next generation neutrino experiment}, \emph{Nuclear Instruments and Methods in Physics Research. Section A, Accelerators, Spectrometers, Detectors and Associated Equipment}, {\bf 695} (2012) pg. 113 $-$ 117.
\bibitem{6}
F.~P.~An et al., \emph{Observation of Electron-Antineutrino Disappearance at Daya Bay
}, \emph{Physical Review Letters}, {\bf 108} (2012) 171803.
\bibitem{7}
D. Naumov, \emph{Shielding EMF at JUNO}, In \emph{JUNO collaboration meeting at Xiamen,China January 12, 2017, JUNO-doc-1300-v1}
\bibitem{8}
V. Baturin et al., \emph{Dynamic magnetic shield for the CLAS12 central TOF detector photomultiplier tubes}, \emph{Nucl. Instrum. Methods Phys. Res. A} {\bf 664} (2012) pg. 11 - 21.
\bibitem{9}
JUNO collaboration, N. Morozov, \emph{simulation of the Earth magnetic field compensation and shielding}, In \emph{JUNO collaboration meeting at Guangdong (Document no. JUNO-doc-260-v1)}, January 13, 2014.
\bibitem{10}
Los Alamos National Lab., \emph{Reference manual for the POISSON/SUPERFISH Group of Codes}, (1987).
\bibitem{11}
H. Lu, E. Baussan and {JUNO experiment}, \emph{The design of the {JUNO} veto system}, \emph{J. Phys. Conf. Ser}, 2017.
\bibitem{12}
http://iaga-aiga.org/
\bibitem{13}
K. Nishijima, \emph{The Super-Kamiokande experiment}, \emph{Radiation Physics and Chemistry}, {\bf 61} (2001) pg. 247 $-$ 253.
\bibitem{14}
W. H. Press, S. A. Teukolsky, W. T. Vetterling and B. P. Flannery, \emph{Numerical Recipes in Fortran 77: the Art of Scientific Computing. Second Edition}, \emph{Cambridge: Cambridge University Press}, 1996.
\bibitem{15}
K. J. Millman and M. Aivazis, \emph{Python for Scientists and Engineers}, \emph{CiSE} {\bf 13} (2011) pg. 9 $-$ 12.
\bibitem{16}
T. Williams et al., \emph{Gnuplot: an interactive plotting program}, \url{http://gnuplot.sourceforge.net/}.
\bibitem{17}
S. Maus, S. Macmillan, S. McLean, B. Hamilton, A. Thomson, M. Nair and C. Rollins, \emph{The US/UK World Magnetic Model for 2010-2015},
British Geological Survey (2010).
\bibitem{18}
G.-Q Zhang et al., \emph{Design of superconducting magnets for self-shielded magnetic resonance imaging based on 0-1 integer linear planning}, \emph{Acta Physica Sinica}, {\bf 61} (2012) pg. 228701.
\end{thebibliography}
\end{document}